\documentclass[a4paper,11pt]{article}
\usepackage{pos}

\title{Prospects for new glueballs and exotics searches}
\ShortTitle{glueballs and exotics}
\author*[a]{Jean-Marie Fr\`{e}re  }
\affiliation[a]{Theoretical Physics CP225, Universit\'{e} Libre de Bruxelles,\\
  blvd du Triomphe , 1050 Brussels, Belgium, \\ and Brout-Englert-Lema\^itre Center, Brussels}

\emailAdd{frere@ulb.be}

\abstract{Glueballs and Hybrids are solid predictions of QCD, but none have this far been identified in an undisputable way. We list several strategies, including the very promising search for "cascade" decays of glueballs and hybrids into each others, and mention the  yet under-exploited sources in heavy ion collisions}

\FullConference{Proceedings of the Corfu Summer Institute 2024 "School and Workshops on Elementary Particle Physics and Gravity" (CORFU2024)\
12 - 26 May, and 25 August - 27 September, 2024\\
Corfu, Greece\\}

\begin{document}
\maketitle

\section{Introduction and recent progress}
While they are the most straightforward prediction of Quantum Chromodynamics (QCD), and expected to be well within reach of even modest colliders , glueballs have not been identified or characterized in a fully convincing way.

Continuing searches have assembled an impressive amount of data, but it is only recently that comprehensive global analysis have been performed, and even now, it is difficult for partial wave fits to have full access to all experiment data.

Traditional approaches based on concepts such as flavor blindness, chiral suppression, or flavor symmetry can fail due to the importance of quantum anomalies on one hand, and to the questionable identification of the resonances in a context of wide states with heavily distorted phase space and overlaps (for instance, recent coupled channel analysis tend to reduce the number of distinct states, which also modifies the assumed branching ratios).

Beyond the direct hadronic production (think fixed targets or hadron colliders),the main gluon-rich source is currently  the radiative decay of heavy vector mesons ($J/\psi)$ for instance, and the BESSIII experiment now has reached the spectral precision to be a major contributor.

A still fledging production source is currently investigated in heavy ion experiments (Alice), but at this stage the analysis is more oriented to the use of possible candidates as an indication of a produced  structure than in the understanding of the glueballs.

The "exotics" (mesons states whose spin-parity-charge combinations cannot be constructed from a single quark-antiquark pair) can shed some light if they are indeed composed of quark-antiquark-gluon.

\section{Simple ideas confront quantum and analysis complications}

The central idea in the attempt to pinpoint glueballs (but which does of course not apply to exotics) is to leverage their blindness to "flavor". While perfectly grounded in theory, it suffers however from serious possible corrections.

The most naive one is "wave-function overlap", which means that in similar decays (say 2 light mesons) the states with the largest overlap with the (comparatively heavy) glueball would be favored (say, 2 kaon final state favored over 2 pions). This is however difficult to quantify a-priori, specially in absence of an explicit decay mechanism.

The assumption of flavor blindness also assumes that final states only connect to glueballs through their quarks, and fails if a direct connection to glue enters the game. There is as strong suggestion that such a link exists through the "strong anomaly", which links the divergence of the axial current (think $\eta$), $\eta'$) to a gluon condensate.
Let us see this in a little more detail.

Consider the divergence of the SU(3) flavor octet and singlet neutral currents,
In absence of anomalies, neglecting also mixing they would be associated to pseudo-goldstone mesons, which would be called respectively $\eta_8, \eta_0$.

\begin{eqnarray}
\partial_\mu A_8^\mu & = & \frac{2}{\sqrt{6}}\,\left( m_u \bar u i \gamma_5 u
+ m_d \bar d i \gamma_5 d - 2 m_s \bar s i \gamma_5 s\right),\nonumber\\
\partial_\mu A_0^\mu & = & \frac{2}{\sqrt{3}}\,\left( m_u \bar u i \gamma_5 u
+ m_d \bar d i \gamma_5 d + m_s \bar s i \gamma_5 s\right)+ \frac{1}{\sqrt{3}}
\,\frac{3}{4}\frac{\alpha_s}{\pi}\,G^A_{\mu\nu}\widetilde{G}^{A\mu\nu},
\label{eq:divergences}
\end{eqnarray}

where $G^A_{\mu\nu}$ is the gluonic field-strength tensor and
$\widetilde{G}^A_{\mu\nu}=\frac{1}{2}\epsilon_{\mu\nu\rho\sigma}
G^{A\rho\sigma}$ its dual.

The usual argument of chiral suppression (which means that a quark-antiquark pair
in a pseudoscalar situation can only couple to 2 gluons at the price of 1 mass factor
for a chirality flip ) would then suggest that (neglecting the u and d quark masses) the $\eta_8$ would couple with a $\sqrt(2)$ factor as strongly as the $\eta_0$.

The physical states are however mixings of the current ones, in a complicated pattern, with the heavier $\eta'$ closer to the singlet state.(see for instance \cite{Akhoury:1987ed}, \cite{Ball:1995zv}, \cite{Escribano:2005qq}.

\begin{figure}[h]
\begin{center}
\includegraphics [width=9cm]{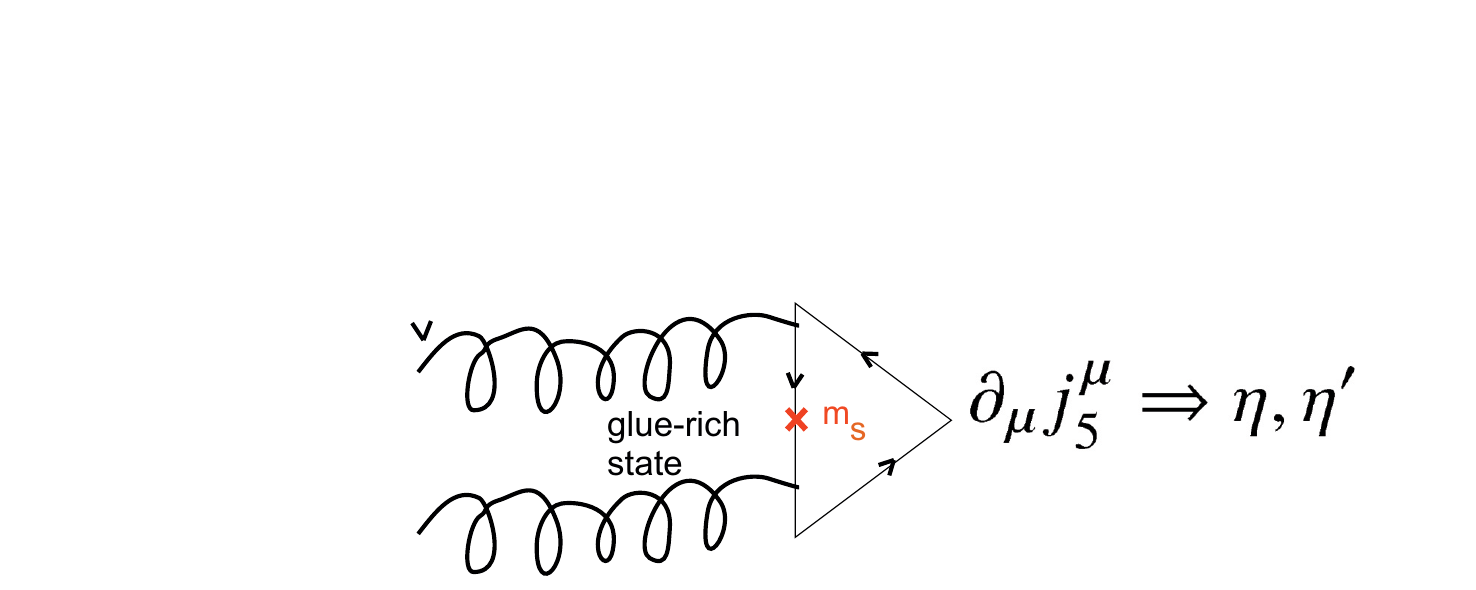}
\caption{Chiral suppression if anomalies are neglected}
\label{GluonstoEta}
\end{center}
\end{figure}

Keeping in mind the anomaly, we see however that the $\eta_0$ connects directly to a pure (pseudoscalar) glue combination!

This was noted \cite{Gherstein} as a possible explanation of the large radiative decay of the $J/\psi$ into $\eta'$ vs $\eta$ (despite the much reduced phase space) \ ( (5.25 vs 1.)  $10^{-3}$).

\begin{figure}[h]
\begin{center}
\includegraphics [width=4cm]{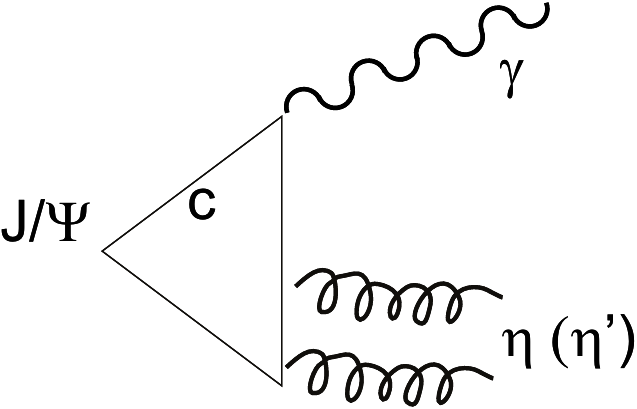}
\caption{a direct connection from glue-rich to $\eta_0$}
\label{JtoGluons}
\end{center}
\end{figure}

We will argue in the following sections that this link between quantum anomaly and glue is a key not only to the glueballs search, but also for the "exotics".

\section{A detour via the exotic mesons}

As we will discuss later, actual glueballs are difficult to identify notably due to the
absence of really clear-cut predictions on the decay branching fractions.

Glue as a "valence" component could be easier to spot in "hybrid" mesons, and offer a key to the later identification of pure glue states.
In particular, "exotic" combinations of quarks and gluons are those with quantum numbers unreachable for pure quark-antiquark mesons, for instance the $\pi_1$, with (I=1, $J^{P,C}= 1^{-,+}$) or the $\eta_1$ with (I=0, $J^{P,C}= 1^{-,+}$).

Even this is not a foolproof way to identify valence gluons, as in "tetraquark" states - 2 quarks and 2 antiquarks - one quark-antiquark pair could mimic the quantum numbers of a gluon, but finding a cluster of related  tetraquarks could then probably help to discriminate the 2 situations.

Candidates for the states $\pi_1$ and $\eta_1$ have been found for a long time \cite{IHEP-Brussels-LosAlamos-AnnecyLAPP:1988iqi}.
For the $\pi_1$, the Particle data Group used to list  2 states, $\pi_1 (1400) , \pi_1(1600)$, but at the time of writing these proceedings, they have essentially been fused. Indeed, a detailed coupled channel analysis (in the pp and pion proton production, including the COMPASS results) has led to this conclusion. \cite{Kopf:2020yoa}

Assuming that the duplication of the states is now a solved question, we must still comment on the decay  modes.
This candidate has been seen from very early on in its $\eta(') \pi$ decay mode, at a time when the expectation was for this decay to be ignored in favor of the $\rho \pi$ mode.

This expectation can be traced to a pioneering paper \cite{DeViron:1984svx}, which unfortunately did not include the quantum anomalies.

We were aware of the GAMS data when we re-examined \cite{Frere:1988ac} the question and showed that the quantum anomaly provided for the $\eta(') \pi$ decay. It is however difficult (in particular in the context of QCD sum rules) to compare in absolute terms the branching ratios into the $\eta$ and $\eta'$ mode.

\begin{figure}[h]
\begin{center}
\includegraphics [width=6cm]{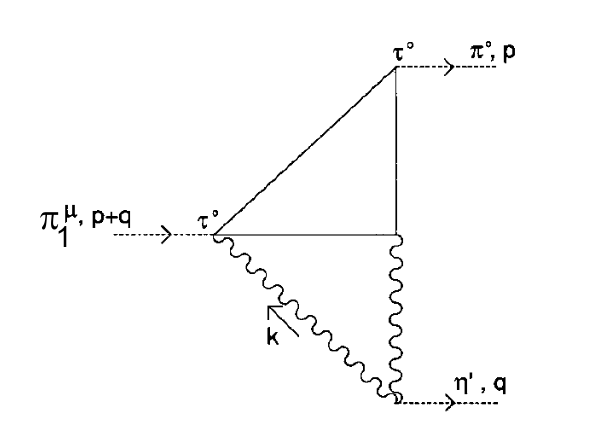}
\caption{a graph allowing the decay of exotics to  $\eta(')$}
\label{Titard}
\end{center}
\end{figure}

Quite interestingly, the ratio of the $\frac{\Gamma (\pi \eta')}{\Gamma( \pi \eta')} \simeq 5.5$ obtained (with a large error bar) by ref.\cite{Kopf:2020yoa} matches the expectations of an anomaly-mediated process. It would be nice to compare to the above-mentioned ratio of $J/\psi$ decays, also in a P wave, but the phase space comparison is hampered here by the large width of the hybrid candidate. Still, the enhancement is quite significative (and pleads for an improved measurement).

This work \cite{Frere:1988ac} was long ignored, but the observaiion by BESIII of the possible exotic $\eta_1$  also presenting the desintegration $$\eta_1 \rightarrow \eta' \eta $$ has brought a parallel analysis \cite{Chen:2022qpd} which we see as a confirmation of our approach.

In fact, like the $\pi_1$; the $\eta_1$ was orignally seen by the GAMS collaboration, but lacked later corroboration.

Why does it turn out (here in the hybrids, but later in the glueballs) that the $\eta(')$ modes are often the discovery channel (and sometimes the only one seen)?
One possibility is of course a strong enhancement (this will have to wait for a better understanding, and a comprehensive analysis of the decay modes), but there could also be a more experimental reason, namely a better signal/noise ratio.
In a way, the $\eta(')$ modes stand out by their heavier mass (in a world heavily laden with pions, in the case of hadronic production) and by their electromagnetic decay , while the comparable mass $\rho$ ends up in the pions background.
Note also that the original GAMS experiments were "protected" by their initial design (a glass wall detecting photons, and thus focused on neutral particles) largely blind to the $\rho$ meson (which does not decay into $ 2 \ \pi^0 $ and thus does not leave a photon mark in the detector). This screening of the $\rho$ modes may have been a great way to reduce background in GAMS!

The $\rho \pi$ modes are indeed very difficult to tease out from the background \cite{COMPASS:2021ogp}.

\textbf{
One interesting direction would be to plead for a comprehensive (but currently lacking) coupled channel analysis involving both the hadronic production and the radiative decays of charmonium.
}

\section{Back to the "bland" Glueball candidates}

At the difference of the hybrids, the glueball candidates have very unspectacular quantum numbers.
The lightest one is expected to have just the "vacuum" quantum numbers, $O^{++}$, while we have already met with the $O^{-+}$ gluon condensate $G^{a \mu \nu} \widetilde{G_{a \mu \nu} }$ through the anomaly graph.

Let us first remark that we deal here also with a very distorted phase space, and that the spectral lines are in general far from simple, with different masses frequently observed in different channels.
Here also, a new tentative coupled channel analysis may shed some light\cite{Sarantsev:2021ein}.
\begin{figure}[h]
\begin{center}
\includegraphics [width=6cm]{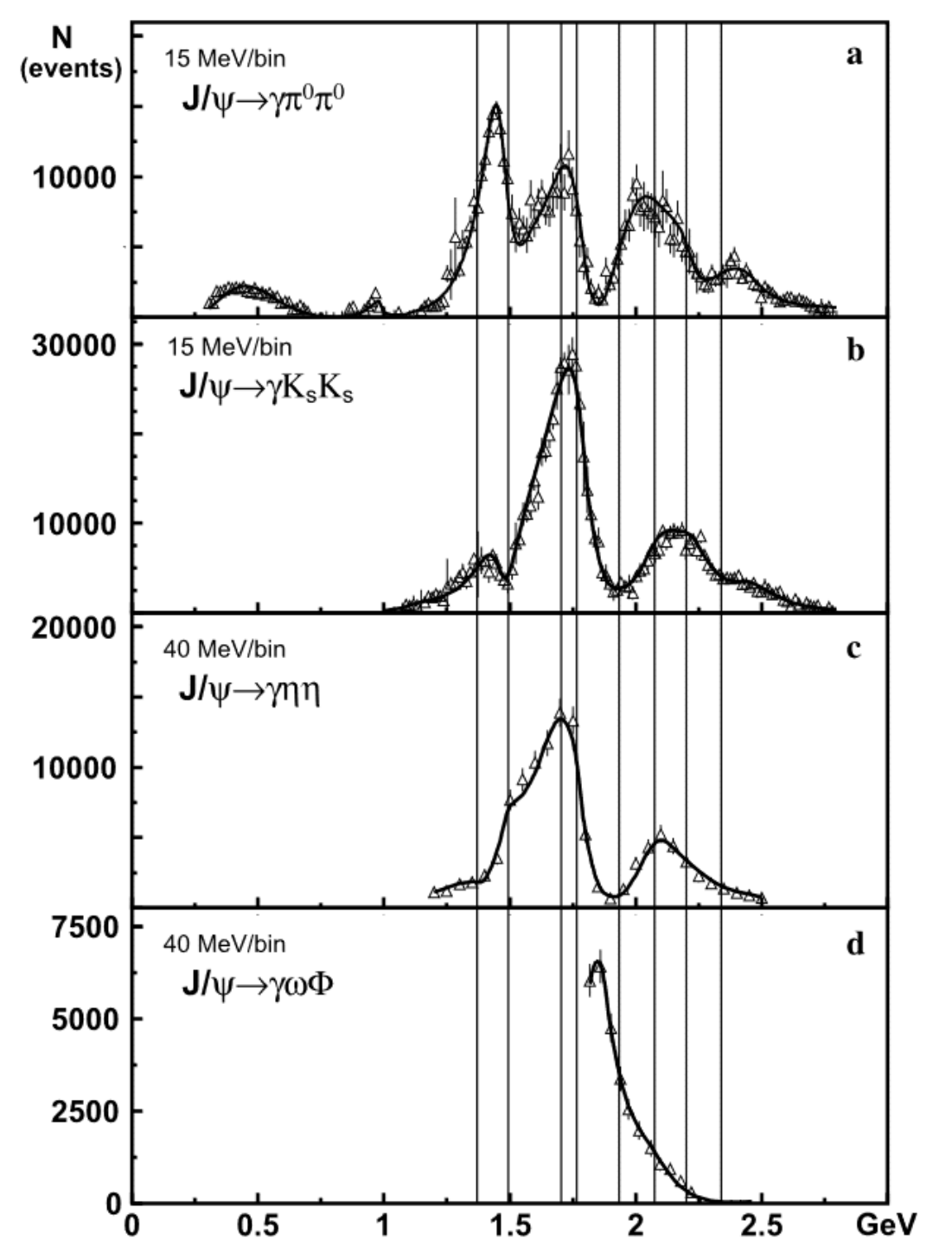}
\caption{Constructive and destructive interference in the pion and kaon modes, from \cite{Sarantsev:2021ein} )}
\label{Sarantsev}
\end{center}
\end{figure}

In this analysis, the authors exhibit notably the presence of an interference in the 2 pions and 2 kaons modes, which they attribute to the sign flip between the strange component of an SU(3) octed and singlet, which are then attributed to the two light modes. (see fig.\ref {Sarantsev}) They argue that a large bump at higher energy corresponding to the hypothetical glueball (in this analysis, the mixing of the light mesons ignores the anomaly, and the glue is treated as a different state).

In the traditional analysis, for  $f_0, 0^{++}$ states, three candidates were usually considered, $f_0(1370)$, $f_0(1500)$ and $f_0(1710)$ respectively in PDG classification (the "mass" quoted here should be seen rather as a label, as the values have differed quite wildly in various observations) .
Various analysis of the data have been performed. They could not be performed from first principles, but relied either on naive SU(3) flavour content, or on supposed graph topologies to relate decay channels (we are a far cry from Feynman diagrams here, no momentum dependance, no phase relations). Depending on the case, direct gluon coupling to the pseudoscalar mesons was included or not, but still in an ad-hoc manner.
For an example and a review of early approaches, see \cite{Frere:2015xxa}.
The recent tendency seems to move from the $f_0 (1500))$ to the $f_0 (1710))$ as the main glueball component.

\begin{figure}[h]
\begin{center}
\includegraphics [width=8cm]{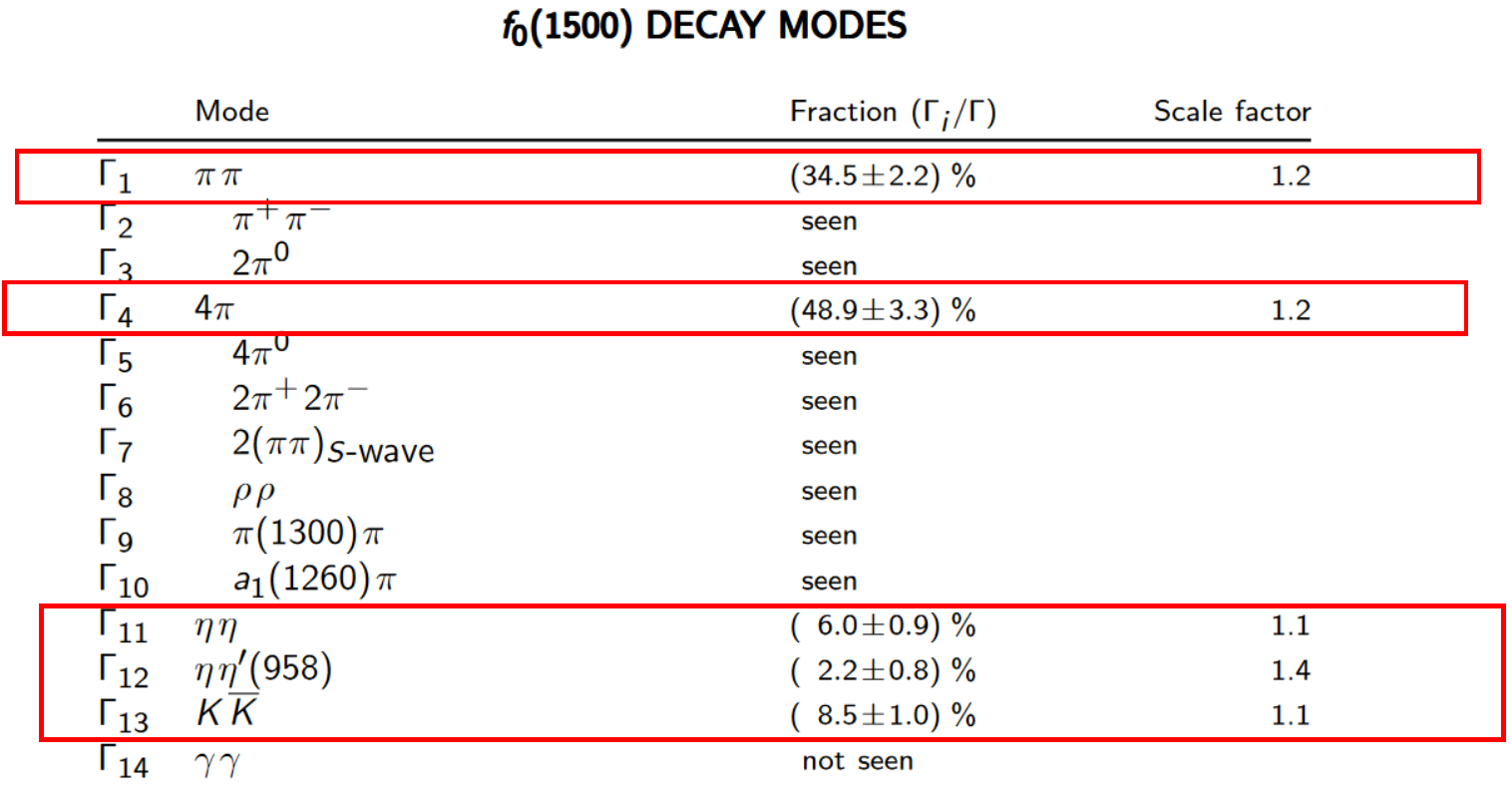}
\caption{$f_0$ decay modes, (from PDG)  )}
\label{f0}
\end{center}
\end{figure}

We still find it interesting (in particular in realtion the the chain decays approach below) to pay some attention to the $f_0 (1500))$ to show its peculiar decay modes.
(see fig. \ref{f0})

The 2 pion modes is of course very present and has the largest phase space, but it is superseded by the 4 pions mode!
It was previously shown that this decay is mainly into  2$\sigma]$ where $\sigma$ stands for $f_0(500)$, a very wide state decaying mainly into 2 pions. This is may be seen as a suggestion of a chain of $O^{++}$ states (more about this in the next section)

The other observation is that the $\eta$  decays are close to the kaon ones, but that, if corrected for  a very small phase space, the $\eta' \eta $ coupling dominates; this once again evokes the preferential coupling of the $\eta$ modes to the gluon component.

\section {Looking for chains of decays}

Another approach, which we think will develop, is to look for chains of decays, assuming that states with a large glueball (or hybrid) component may prefer to decay in the same vein. We have already seen this above with the $f_0 (500)$ and the
 $f_0 (1500)$. \\

\textbf{In an important and more significant  example, }BESIII observed a possible pseudoscalar glueball candidate $X2600$ \cite{BESIII:2022sfx} (for a review, of this topic see \cite{Chen:2022asf}).
This review mentions a "strong correlation" between the pseudoscalar glueball candidates and the $f_0 (1500)$ , with  $B(J/\psi\rightarrow\gamma X(2600))\cdot
B(X(2600)\rightarrow f_{0}(1500)\eta')\cdot
B(f_{0}(1500)\rightarrow\pi^{+}\pi^{-})$
=
($3.39\pm0.18^{+0.91}_{-0.66})\times10^{-5}$

A similar branching is observed for the tensor glueball candidate $f_2'(1525)$, a tensor glueball candidate.

\textbf{We hope to see here the first cases of glueball characterization through a chain of preferential decays. }

It is interesting to put the above result in parallel with similar measurements for the lower states
We extract the following table from \cite{BESIII:2022iwi}

Of course the branching ratio may seem small, but this is in fact due largely to the high number of decay modes of he $J\psi$

\begin{figure}[h]
\begin{center}
\includegraphics [width=12cm]{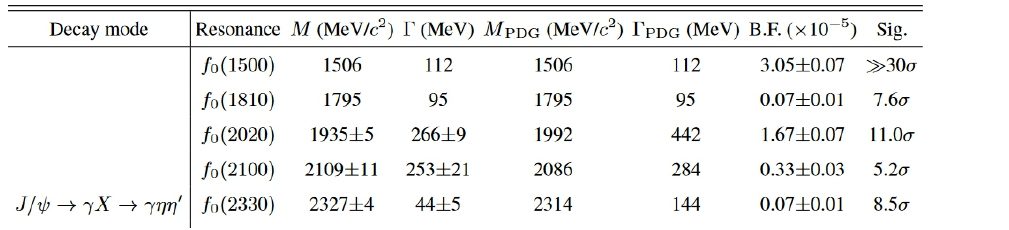}
\caption{possible glueballs in radiative  $J/\psi$ decay as a glue-rich source coupled to the $\eta \eta'$, The product of the branchings is a possible indicator of glueballs; this is part of a table from \cite{BESIII:2022iwi} }
\label{BESiiiTable}
\end{center}
\end{figure}

We note that in this chain, the $f_0 (1500)$ and the complex of states above 2 GeV are dominant.

In fact this could be a very interesting approach  \textbf{ the combined branching ratio} $J/\psi \rightarrow \gamma X , X \rightarrow final state$ \textbf{which effectively quantifies the product of "glue-rich production" and of "glue-rich decay".}

A side remark about branching fraction: we don't have in general a precise estimation of the width of the states (highly distorted phase space, coupled channel interference), and thus of the absolute branching fraction. We can't in fact count on an enumeration of the decay modes, as some are either not searched for (think of multiple pions, where we know that for $f_0 (1500)$ the 4 pions mode is larger than the 2 pions, but the n>4 modes are not looked for) , or impossible to extract from the background.

\textbf{We also remark here the recurrent presence of the $\eta'$ in all the glueballs and hybrid states searches.}

\section {Heavy ions}

\begin{figure}[h]
\begin{center}
\includegraphics [width=6cm]{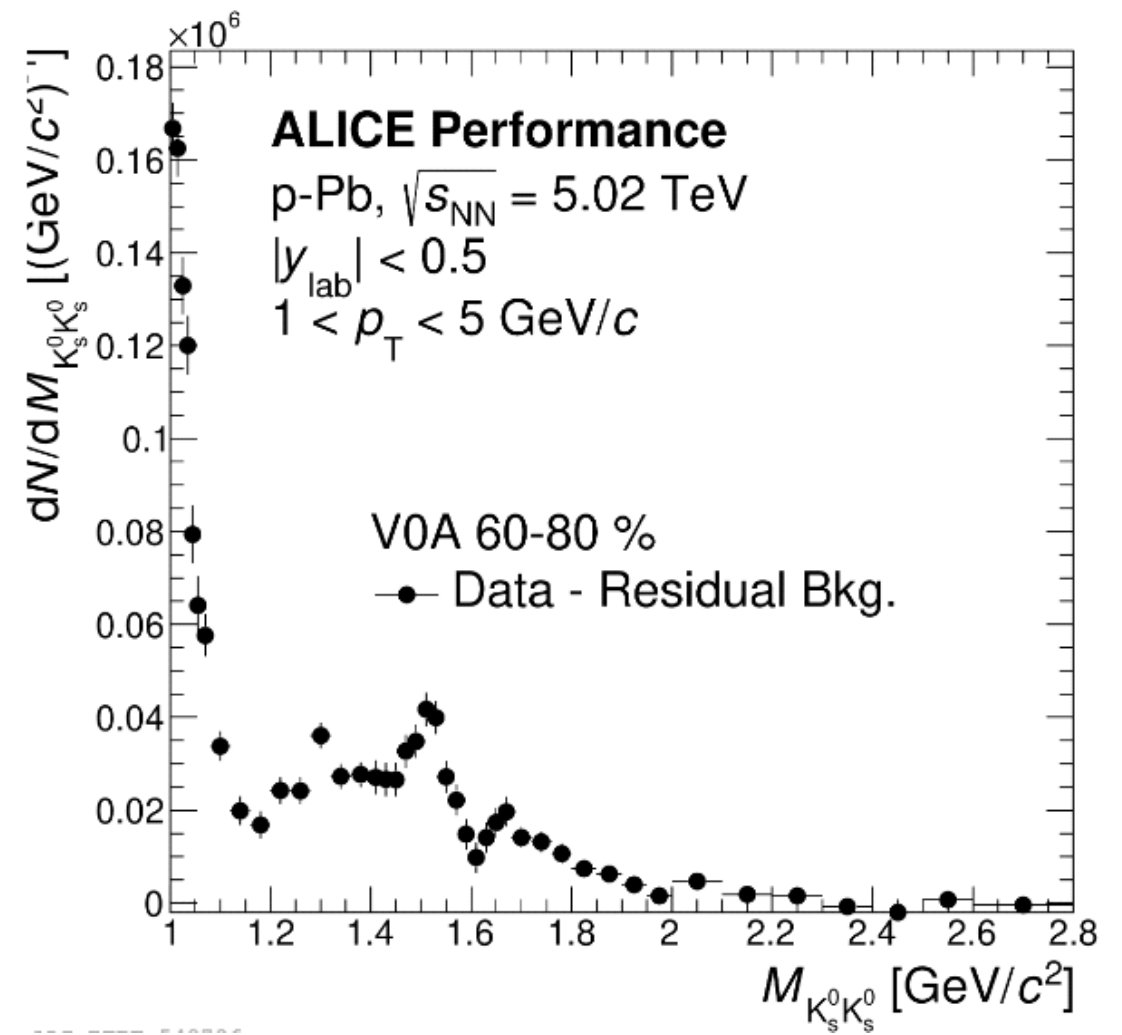}
\caption{Hints for glueballs in Alice collisions)}
\label{Alice}
\end{center}
\end{figure}
We need to make here a special mention of the heavy ion collisions. For the moment, it seems
that glueball candidates are used here merely as a way to characterize the collision.
They are more likely to be identified by mass than by decay mode.
We are in early stages here, and heavy ions certainly provide for "glue-rich" situations.
Progress is certainly possible, with glueballs helping to understand the collision, and
the production characteristics bringing information for the identification of glueballs.
For lack of more detailed results, we paste here from an Alice Poster in Quark Matter 2023. (fig \ref{Alice})

\section {Prospects}
This far, glue-related states have been looked for in several sources,

\begin{itemize}
  \item central region of hadronic collisions
  \item proton-antiproton annihilation
  \item radiative decay of the $J/\psi$ (in particular at BESIII
  \item other quarkonium decays (B mesons)
  \item very recently, in heavy ion collisions
  \item and progressively, our hopes turn to "glue" cascade chains from heavier glueball candidates
\end{itemize}

\end{document}